# Design of an IF section for C band polarimetry

Miguel Bergano, Luis Cupido, Domingos Barbosa, Rui Fonseca, Bruce Grossan, George Smoot

*Abstract* — In the context of the Galactic Emission Mapping, a new receiver at 5GHz was developed to characterize the galactic foreground to the Cosmic Microwave Background Radiation. This is a 5GHz super heterodyne polarimeter with double down conversion, with a high gain IF chain using the latest RF technology working at 600MHz central frequency that feeds a four channel digital correlator. This paper describes the receiver and its current status. Design options and constraints are presented with some simulations and experimental results of a circuit prototype.

## I. INTRODUCTION

THE Cosmic Microwave Background Radiation (CMBR), with its small temperature fluctuations (50~80 μK) [6] and its small polarization variations (1~10 μK) [7] is the best cosmological probe of the early Universe, allowing a reconstitution of the cosmic history with great precision. However, CMBR is partially obscured by radiation emitted by our galaxy, the Milky Way. While a lot of knowledge has been acquired in recent years on the total flux of galactic emissions, a lack of knowledge on the sky polarized emissions has prompted the need for detailed maps of the polarized galactic foreground emissions to properly extract the information from CMBR observations. The GEM (Galactic Emission Mapping) collaboration [3, 4, 5] started with the aim of mapping the sky with high sensitivity and absolute calibration at low frequencies both in all sky [1]. For the Northern Sky Hemisphere, a GEM team will proceed towards a survey of the polarized galactic emission at 5GHz with a 9 meter Cassegrain antenna installed in central Portugal. Although the radiation from the galaxy is considerably stronger than the CMBR at 5GHz, it still requires a high sensitivity radiometer/polarimeter, with an equivalent noise temperature around 10K. For this purpose a heterodyne radiometer/polarimeter was being developed, using cryogenically InP preamplifiers and 600MHz IF (intermediate frequency) followed by an all-digital correlator built using an FPGA (field programmable gate array).
The present article details the IF section of the receiver and the aspects of IO to the digital correlator. The cryogenically cooled front-ends will be published elsewhere.

## II. RECEIVER OVERVIEW

The receiver has a general classical heterodyne topology, the back-end being fully implemented in the digital domain (Figure 1). In radiometry the bandwidth should be as large as possible to permit the best instrument resolution. This is however limited by the available bandwidth free of interference and preferably under protection of the international frequency allocations for the RA (radio-astronomy) service. In our case we want a minimum bandwidth of 200MHz around 5GHz. However, the center frequency had to be changed to 4.9GHz (using the same 200MHz bandwidth) in order to be aligned with the band segment allocated to RA for which we can apply for protection to the Portuguese radio spectrum administration (ANACOM). The radiometer/polarimeter receiver is a double conversion super-heterodyne receiver with zero-IF. The front-end will use cryogenically cooled InP HEMTS preamplifiers from the Low Noise Factory in Chalmers, Sweden. It is followed by a 1-stage low noise amplifier using an AVAGO pHEMT transistor at room temperature (ATF360-77 chip) to add gain and decrease the overall noise contribution to the receiver from the filter and mixer-LO circuitry. Performance of this chip will be described elsewhere (in preparation). Next is the first selectivity stage of the system an image rejection filter succeeded by diode mixers along with a local oscillator (LO). All this equipment will be located at the back end of the antenna feed inside a temperature shield – RF boxes. The remaining part of the IF chain is composed of a preamplifier-filter and a large gain IF amplifier followed by a converter to zero-IF that provides the base-band signals for the digital correlator. The Base-band signals are digitized and fed to an FPGA where the actual correlation process happens, all done entirely in the digital domain. Signal digitization is performed at 200Ms/s (Mega Samples per Second) with 8 bits of resolution. For this application the required resolution at the ADC is quite low, since the number of samples that are integrated while acquiring a pixel is extremely high (8ns per sample over a few ms of temporal resolution results in a 1:100000 ratio where amplitude and phase resolution came essentially from the integrated signal statistics rather than ADC resolution). Therefore the market availability of 8 bit ADCs dictated this choice. The correlated samples from which the Stokes Parameters (I, Q, V and U) are calculated will determine the polarization gathered by the antenna. These parameters are integrated in the FPGA and the results sent to a PC. To avoid mechanical moving parts a small form factor industrial PC104 is used. A paper related to the implementation in the FPGA of the Stokes parameters calculation will be published elsewhere, along with the description of the communication between the PC104. The

·GEM Portugal team acknowledges Portuguese Foundation for Science and Technology (FCT) for the financial support through POCI/CTE-AST/57209/2004 and PTDC/CTE-AST/65925/2006 project grants. DB is supported through a Ciência 2007 grant, funded by COMPETE and QREN. RF was supported with a BD fellowship from FCT.

M. B. is with the Radio astronomy Group of Institute of Telecommunications, Aveiro, Portugal (phone: 351-234377900; fax: 351-234377901; e-mail: jbergano@av.it.pt).
L. C. is with the Instituto de Plasmas e Fusão Nuclear, Instituto Superior Técnico, Portugal
D.B. and R.F. are with the Radio astronomy Group of Institute of Telecommunications, Aveiro, Portugal
B. G. and G.S. are with the Lawrence Berkeley National Laboratory, University of California, at Berkeley, USA

expected overall system temperature is about 10K corresponding to a total power (over 200MHz of bandwidth) of about -105.6dBm therefore the receiver would need a total gain of about 104dB on the RF and IF altogether to provide the necessary signal level for digitalization. The gain budget analysis for the system is shown on Table I, corresponding to the block diagram shown on Fig. 1.

TABLE I
POWER BUDGET FOR THE COMPLETE RADIOMETER / POLARIMETER

| Antenna | LNA | 2$^{nd}$ LNA | Passive Filter | Mixer | IF amplifiers | Converter | ADC |
|---|---|---|---|---|---|---|---|
| Input (dBm) | 36 | 13 | -4 | -7 | 64 | 2 | Output (dBm) |
| -105,6 | -70 | -57 | -60,6 | -68 | -3,6 | -1,6 | -2 |

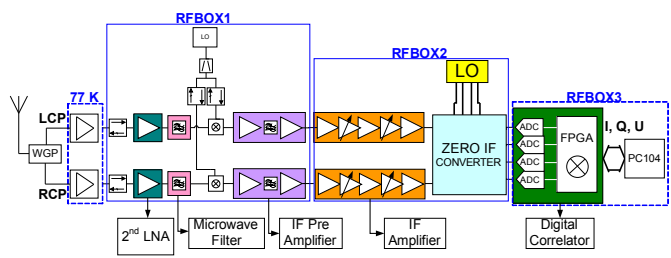

Fig. 1 Overall block diagram

Another important issue while designing a very high sensitivity receiver is the noise figure of each element or the equivalent temperature that each element contributes to the system. In order to do such analysis there is in Table II the noise budget of the initial blocks, the main contributors.

TABLE II
NOISE BUDGET

|  | LNA | 2$^{nd}$ LNA | Filter + Mixer + IF Pre Amplifier |
|---|---|---|---|
| $T_{eq}$ | 6K | 36K | 6202K |
| NF | 0,325 dB | 0,5 dB | 13,5 dB |
| Gain | 36 dB | 13 dB | 20 dB |
| Linear Gain | 3891 | 20 | 100 |

In the previous Table we neglect for purpose of clarity here the contribution of the cables and all the SMA connectors. It can be verified in the previous table that without the second LNA the overall noise figure of the system will increase due to the contribution of the Filter+Mixer+IF Amplifiers noise figure. Using another LNA will reduce this contribution.

### A. IF Pre amplifier

The signals coming from the front-end block (located at the feed point) enter the IF unit of the receiver directly to the IF preamplifier and filter module. This module presents a gain of 31 dB with a moderately low noise figure. The design was targeted for gain and proper IF bandwidth shaping purposes but other considerations were taken into account, such as gain variation with temperature and frequency.

The gain stability required relates directly to the kind of data that is going to be extracted from the received signals. In order to observe sub-miliKelvin of antenna temperature, we need to ensure that gain variations would preferably be in a different time scale and smaller than the measured variations. The variation of gain with environmental temperature needs to be minimized at all cost. Therefore amplifiers with minimum gain variation with temperature were selected.

Encapsulated MMIC (monolithic microwave integrated circuits) were attractive choices and their possible use was investigated. As shown in Figure 1 this module has two amplifier stages with a band-pass filter between them. The amplifiers used were ERA3 from Minicircuits due to their small temperature drift in the frequency band of interest (500MHz to 700MHz). The manufacturer claims a gain increase with temperature of 0.12 dB from -45 to 85ºC, which corresponds to 0.0009dB/ºC. The filter section is a Butterworth type with 200MHz bandwidth centered at 600MHz. It is implemented in a T configuration with lumped L and C elements. To allow calibration the filter uses trimmer capacitors and fixed inductors. The design of the filter as well all the other parts was computer aided, using the ADS (Advanced Design System) software [23]. A fine adjustment was made in the final design by trial and error, in order to achieve the desired bandwidth and response flatness with commercially available component values. Very wideband MMIC's do not exhibit a constant gain over frequency, higher frequencies presenting the lowest gain. In order to reduce this effect we inserted a slope compensation network, which is a simple RLC network, next to the filter that reduces the gain at lower frequencies by loading the circuit more heavily at lower frequencies. On the modeling we used ERA3 and ERA2, both from Minicircuits, and the results are shown in Fig. 2.

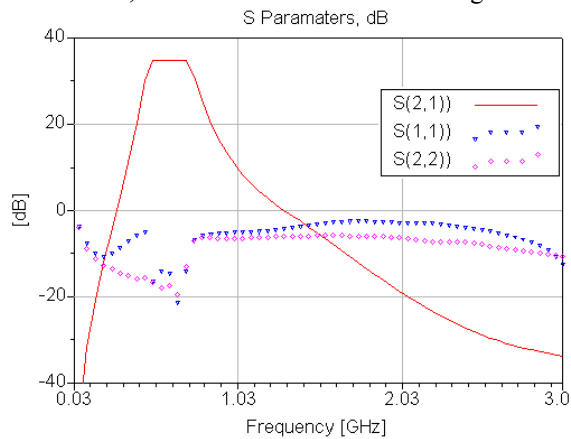

Fig. 2 IF preamplifier simulation on ADS2005.

The layout uses microstrip lines, and the frequencies involved suggested the usual RF PCB design techniques with 0805 or 0603 size SMD components. The PCB board is packaged in an aluminum milled box specially designed for this type of application to serve both as shielding and thermal mass. The tests were made using an HP 8753E Network Analyzer and results are in Fig. 3.

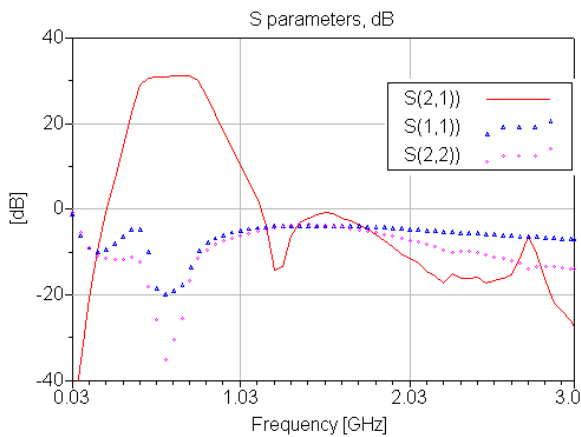

Fig. 3  IF preamplifier measurement results as measured on a HP8753E network analyzer.

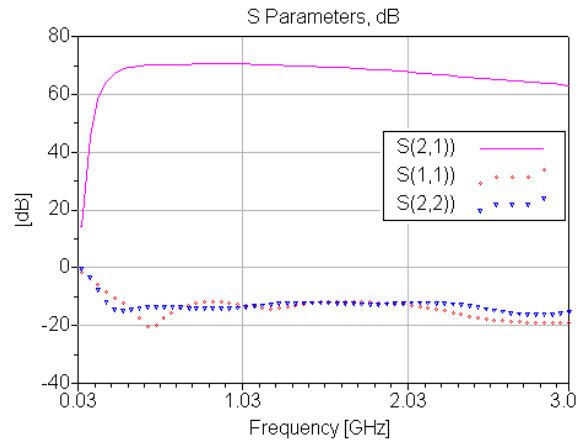

Fig. 4  IF amplifier simulation on ADS2005.

### B. IF Amplifier

This stage is the main IF amplifier and will provide most of the gain required. It is composed of five MMIC stages and two digitally controlled attenuators. Considering the values expressed in Table I we would require about 56dB of gain, however, by design, we should operate near the middle of gain control settings which is about 15dB below the maximum gain (for a total gain control of 30dB ). In this way a total gain of 71dB would be necessary. Controlling the amount of signal at the input of the ADCs is easily accomplished by the MMIC digital RF attenuators DAT-15R5-PP Digital step attenuators from Minicircuits. Each of these has an attenuation of 15,5dB and is controlled by 5 bits with a 0,5dB step. The variation of gain with frequency and temperature was also taken into account, and in this case we would benefit from a very flat band from 100MHz to 1GHz allowing us to have a perfect flatness between 500MHz and 700MHz. The MMIC's chosen were one ERA3 (21dB) and four ERA2 (15,5dB). The expected decrease in gain of the MMIC amplifiers at higher frequencies was again corrected by inserting slope compensation networks, like we did for the IF preamplifier. This time we inserted two RLC networks for the entire module. The attenuator was modeled using simple resistive circuits, which proved to be accurate enough for our needs. Simulation results can be seen in Fig. 4, after obtaining S parameters for the ERA3 and ERA2 from the manufacturer.

As for the IF preamplifier microstrip lines were used, and the frequencies involved suggested the usual RF PCB design techniques with 0805 or 0603 size SMD components. Packaging was also approximately the same. The test results, obtained with an HP 8753E Network Analyzer are shown on Fig. 5.

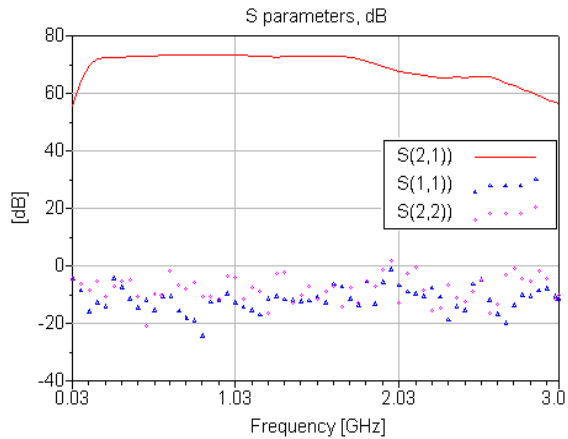

Fig. 5  IF preamplifier results measured on a HP8753E network analyzer.

### C. Converter

The signal from each arm of the IF needs to be converted down to a band of frequencies that allows the signals to be processed in the digital domain. Since we need to preserve 200MHz bandwidth we have two options: i) converting down to 0 to 200MHz and acquire at 400Ms/s (or slightly higher , in order to account for filter roll off and aliasing issues); ii) converting to zero IF, e. g., to -100MHz to +100MHz acquiring at 200Ms/s (or slightly higher, for the same reasons as above) but in order to preserve the total 200MHz bandwidth a complex signal conversion is required with base-band I and Q signals produced for each arm of the receiver (RHCP – Right Hand Circular Polarization – and LHCP – Right Hand Circular Polarization). By using the second option, the complex conversion to a zero-IF, we end up with 4 channels to be digitized but at half the sampling rate that would be required otherwise. This is extremely convenient in order to relax as much as possible the requirements for both the ADC and FPGA. The signals from the two arms of the radiometer are split into its phase (I) and quadrature (Q) components. The converter produces the I and Q outputs by multiplying the IF signal with two versions of the local oscillator with a phase difference of 90º. This operation will be applied to both RHCP and LHCP arms, so the implementation uses two identical circuits

The 600MHz IF signal is separated into two channels,

using a power splitter ADE-2-9 from Minicircuits to feed a pair of ADEX-10 mixers. The local oscillator (LO) already outputs the correct driving level for the mixers, 7dBm, and the correct phase relation, 0 and 90º of phase difference. The I and Q signals obtained will be low pass filtered to eliminate unwanted frequencies, then amplified and finally low-pass filtered prior to digitization. A diagram of the converter is shown on Fig. 6.

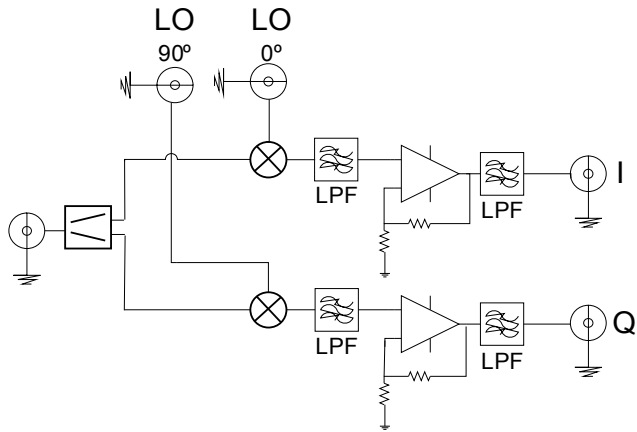

Fig. 6 Zero-IF converter block diagram

The filters that follow the mixers are 7th order Butterworth low-pass filter with 100MHz cut-off frequency. The output filters are 3rd order Butterworth low-pass with 120MHz cut-off frequency. This last filtering prior to the digital acquisition of signals will eliminate any wideband noise and spurious signals that might be presented at this point ensuring that only the signals of interest are presented to the ADCs. Both filters were implemented using lumped L and C elements. The amplifier uses the OPA657, an OPAMP (operational amplifier from Burr-Brown) in a non-inverting configuration. The layout follows the usual techniques for RF instrumentation using both PCB layout and external metallic shielding. Further care was necessary to externally ensure phase balance both in the RF and LO signal path, in order to guarantee perfect orthogonal output signals in all circumstances. Fine phase trimming was provided with a variable capacitor to allow a precise calibration. The circuit employs microstrip layout design and is constructed on a FR4 epoxy substrate. For the tests we used an RF signal of 600 to 700MHz with -7dBm and an LO signal at 600MHz with 7dBm. The output signal, varying from DC to 100MHz was measured with a spectrum analyzer and presented in Fig. 7.

*D. Local Oscillator*

This circuit provides the LO signal to the converter. Since we want to do a zero-IF conversion, the required oscillator frequency is 600MHz, the exact central frequency of the IF pass-band. We decided to use the encapsulated oscillator ROS615 from Minicircuits which is a VCO (voltage controlled oscillator) that tunes from 580 to 615 MHz. It is operated at the fixed frequency of 600.0MHz. We have both the possibility to tune the frequency using a potentiometer, varying the voltage from 0V to 5V or we may use a PLL synthesizer chip LMX2326 from National Semiconductors. As for the moment no frequency stability better than 1MHz is required, so the LO operates with analog frequency control.

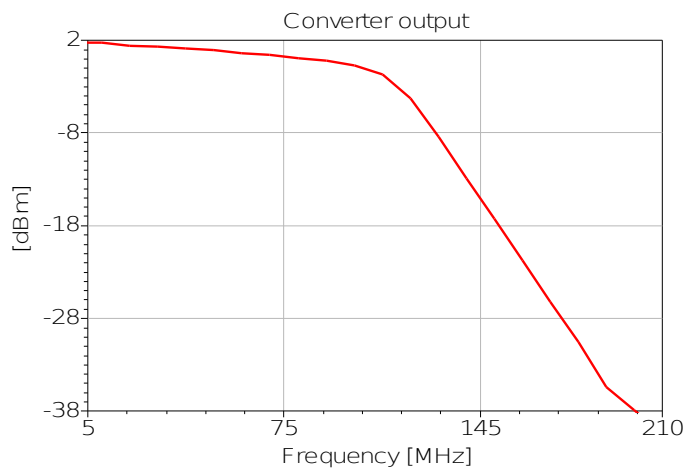

Fig. 7 Converter results as measured on a HP8563A spectrum analyzer.

required, so the LO operates with analog frequency control.

The oscillator signal needs to be split into four equal signals. For simplicity we accomplished this with simple resistive dividers in T attenuator configurations allowing for normalization of the gain and attenuation required, thus providing four identical signals close to 7dBm. Since the signal from the VCO was considerably low (about 0dBm) we needed to amplify it before we split the signal. In order to have isolation between the output ports of this unit it is desirable to have one amplifier per output. Taking these requirements into account we arrive to the design presented on the diagram of Fig. 8.

We used an ERA1 (from Minicircuits) MMIC amplifier to raise the power of the VCO to about 10dBm. The resistive divider then lowers each output power to about -4dBm. For this reason another ERA1 amplifier is needed to raise output power again to 7dBm, the power required to drive the converter block.

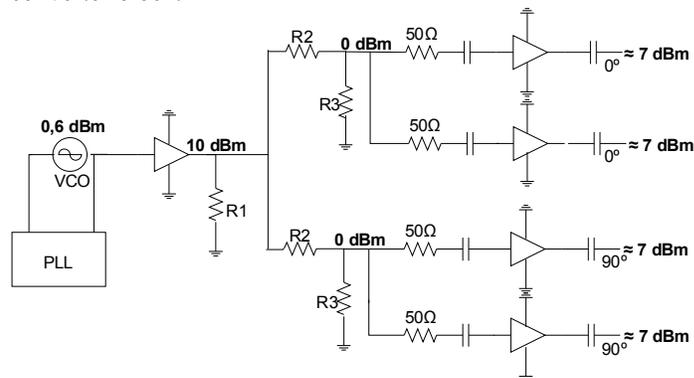

Fig. 8 Local oscillator module block diagram.

The same layout considerations and RF common practices apply for the local oscillator module. This block was also implemented on FR4 substrate using microstrip design and SMD components.

III. CONCLUSIONS

To accomplish the polarimetric measurements at 5GHz for the Galactic Emission Mapping in the North Hemisphere we implemented a heterodyne polarimeter with all final polarimetric measurements being performed in the digital domain. For this task, a new intermediate frequency chain based on a high performance IF strip attended by a digital correlator, suitable for a radio-astronomy application, and in particular for the GEM (at Portugal) experiment, was designed, constructed and tested successfully.

This IF chain will feed the FPGA based correlator capable of outputting the 4 Stokes parameters of the incoming radiation. The entire IF system here described was therefore implemented using standard SMD components, classical approaches to high frequencies and microstrip design, using state-of-the-art commercial RF MMIC devices.

## References

[1] Amici, Giovani De, *GEM Setup Requirements*, Astrophysics Note 478, 1995, Space, Sciences Laboratory of California at Berkeley
[2] Banday, Antony John, Fluctuations in the Cosmic Microwave Background, PhD. Thesis, 1992.
[3] Bensadoun, M., et al., Measurements of the Cosmic Microwave Background Temperature at 1.47 GHz, Ap. J. 409:1-13, 1993.
[4] Bohorquez, Camilo Tello, Um Experimento para Medir o Brilho Total do Céu em Comprimentos de Onda Centimétricos, Tese de Doutorado em Ciência Espacial / Radioastronomia e Física Solar, INPE, 1999.
[5] Bohorquez, Camilo Tello, RFI Monitoring at 2.3 GHz in Brazil, ?, 2001.
[6] Brandt, W. N., Separation of Foreground Radiation from Cosmic Microwave Background Anisotropy using Multifrequency Measurements, Ap. J. 424:1-21, 1994.
[7] Cohen, M. H., Radio Astronomy Polarization Measurements, Proc. IRE, vol. 46, pp 172-183, January, 1958.
[8] Colin, R. E., Foundations for Microwave Engineering, McGraw-Hill, 1966.
[9] Cortiglioni, S., et al, The Limitations of Cosmic-Microwave-Background Measurements due to Linear Polarization of Galactic Radio Emission, Astron. Astrophysics., 302, 1-8, 1995.
[10] G. l. Matthaei, L. Young, and E. M. T. Jones, Microwave Filters, Impedance-Matching Networks, and Coupling Structures, Artech House, Dedham, Mss., 1980.
[11] Skou N., "Microwave Radiometer Systems", Artech House
[12] Marconi G., "Wireless Telegraphic Communication", Nobel Lecture, 1909
[13] Barbosa D., "The relationship of Physics and Telecommunications", ConfTele 2005, Portugal
[14] Wilson, Robert W., "Cosmic Microwave Background Radiation", Nobel Lecture in 1978
[15] Penzias A., "The origin of elements", Nobel lecture in 1978
[16] M Tegmark, "CMB mapping experiments: a designer's guide", , Phys. Rev. D, 56, 4514-4529, 1997
[17] D. Barbosa, José M. Bergano, Rui Fonseca, Dinis M. dos Santos, Luis Cupido, Ana Mourão, George F. Smoot, Camilo Tello, Ivan Soares, Thyrso Villela "The Polarized synchrotron with the Polarized Galactic Emission Mapping Experiment, referred in CMB and Physics of the early Universe", International Conference, in Proceedings of Science, Sissa, PoS(CMB2006)029, 2006
[18] G. Smoot et al., "Structure in the COBE differential microwave radiometer first-year map", Astrophysical Journal, vol. 396, pp. L1-L5, Sept. 1992
[19] L. Page et al, "Three Year Wilkinson Microwave Anisotropy Probe (WMAP) Observations: Polarization Analysis", submitted to Astrophysical Journal, astro-ph/0603450
[20] S. B. Cohn, "Parallel-Coupled Transmission-Line-Resonator Filters", IRE Trans. Microwave Theory and Techniques, vol. MTT-6, pp. 223-231, April 1958.
[21] Graham, M. H., "Radiometer Circuits", Proc. IRE, Vol. 46 1958
[22] Miguel Bergano, Francisco Fernandes, Luis Cupido, Domingos Barbosa, Rui Fonseca, Dinis M. Santos,George Smoot, "C-Band Polarimetry using a full digital correlator", submitted for URSI2007, Tenerife, Setembro de 2007
[23] Agilent Technologies, Advanced Design System 2005A, Agilent EEsof EDA
[24] http://www-astro.lbl.gov/gem/
[25] http://lambda.gsfc.nasa.gov/product/cobe/
[26] http://www.cea.inpe.br/~cosmo/index_gem.htm
[27] http://www.av.it.pt/gem
[28] www.nrao.edu/whatisra/history.shtml

**José M. Bergano** is an Msc in Electronic Engineering by University of Aveiro, where he is currently a PhD student developing work related to the GEM receivers and instrumentation for radio astronomy.

Rui Fonseca is a physics engineer from Instituto Superior Técnico in Lisbon. He is currently doing is Ph.D. in electronic engineering at Instituto de Telecomunicações in Aveiro within the Galactic Emission Project in Portugal.

Luis Cupido is an electronic engineer from U. Aveiro and got its Ph.D. in Physics Instituto Superior Técnico in Lisbon. He is an expert on Radio Astronomy and Microwave reflectometry for Nuclear Fusion projects like JET and ITER. He had developed high performance devices and instruments for microwave and millimeter wave applications, like oscillators, mixers, LNAs and a well-known VCXO Phase-lock loop to GPS. Currently is member of CRAF commission of the ESF for radio spectrum protection.

Domingos Barbosa got its Ph.D. in 1997 from the University of Paris VII in Astrophysics and Space Techniques. He was research fellow at Imperial College and Lawrence Berkeley National Laboratory and Instituto Superior Técnico. He is Ciência 2007 researcher at Instituto de Telecomunicações - Aveiro on Experimental Radio Astronomy. He is the PI of the GEM project in Portugal and collaborated with several projects like MAXIMA and Planck.

Bruce Grossan, is a PhD in Physics from MIT and works on astrophysics and cosmology projects at LBNL, USA, involving the Cosmic Far infra-red Background, AGN, and interstellar medium. He had worked on 10 GHz ground-based CMB measurements. He is currently the Principle Investigator on the Cosmic Far-IR Background projects on the Spitzer Space Telescope, and also works on the GEM project with Dr. Smoot.

George Smoot is a Ph.D. (1970) in Physics from MIT and has been at the University of California Berkeley and the Lawrence Berkeley National Laboratory since 1970. He was the PI of the COBE/DMR satellite experiment (1989-1994) that detected for the first time the CMB anisotropies. He is Nobel Prize in Physics in 2006 for its discoveries with COBE, and was awarded the Albert Einstein Medal (2003), the Gruber Prize (2006) and the NASA Medal for Exceptional Scientific Achievement (1991). He was member or co-I of several projects in Space Sciences, Experimental Cosmology, Astrophysics and High Energy Physics.